\documentclass[a4paper,11pt]{article}
\usepackage{pos}
\usepackage{graphicx,amsmath,amsfonts}
\usepackage{wrapfig}
\newcommand{\inspire}[1]{[\href{https://inspirehep.net/literature?q=#1}{\sc inSPIRE}]}
\usepackage{supertabular}
\usepackage{slashed}
\usepackage{multirow,multicol,diagbox,array} 
\usepackage{slashed}

\allowdisplaybreaks[2]
\usepackage{calrsfs}
\DeclareMathAlphabet{\pazocal}{OMS}{zplm}{m}{n}
\def\pt{\widetilde p}
\def\pa{p_A}
\def\pn{p_N}
\usepackage{color}

\definecolor{mycolor}{rgb}{0.6,0.0,0.4}

\title{Nuclear modifications on longitudinal-transverse structure-function ratio $R$
in the deuteron}
\author[a,b,c]{S. Kumano}
\affiliation[a]{Quark Matter Research Center,
    Institute of Modern Physics, Chinese Academy of Sciences,\\
    Lanzhou, 730000, China}
\affiliation[b]{ Southern Center for Nuclear Science Theory,
    Institute of Modern Physics, Chinese Academy of Sciences,\\
    Huizhou, 516000, China}
\affiliation[c]{KEK Theory Center, Institute of Particle and Nuclear Studies, KEK,
    Oho 1-1, Tsukuba, 305-0801, Japan}
\emailAdd{kumanos@impcas.ac.cn}

\abstract{
In lepton scattering from nuclei, it has been taken for granted that 
nuclear modifications do not exist for the longitudinal-transverse 
structure-function ratio $R$, although nuclear effects in $F_2$ have been investigated
for a long time. In fact, experimental data of lepton-nucleus scattering
have been analyzed with this assumption. It is obviously not appropriate because
nuclear modifications exist in the function $R$ theoretically as shown in this work.
In this work, we explain nucleon's Fermi motion's effects by using a convolution
description for nuclear structure functions, especially on the deuteron.
The longitudinal and transverse structure functions are defined by
taking the virtual photon momentum direction along the $z$ axis.
However, nucleons in a nucleus could move in any direction, so that
nucleon's longitudinal and transverse structure functions could mix
with each other in the nuclear medium. Such mixing 
effects could be of the order of $\vec p_\perp^{\,2}/Q^2$,
where $\vec p_\perp$ is the nucleon's transverse momentum and
$Q^2$ is given by $Q^2=-q^2$ with the virtual-photon momentum $q$,
because the transverse Fermi-motion is the source of such a mixture. 
In addition, nuclear effects are different between the longitudinal 
and transverse structure functions in the convolution model
because their $x$-dependent functional forms are different.
Convolution integrals have different results between the longitudinal 
and transverse structure functions for nuclei.
Because the experiment is in progress at 
the Thomas Jefferson National Accelerator Facility to measure
the function $R$ for the deuteron, the numerical results are shown
on the nuclear effects of $R$ in the deuteron.
Hopefully, such nuclear effects are found experimentally by future
experiments including heavy nuclear targets.
}

\FullConference{33rd International Workshop on Deep Inelastic Scattering 
and Related Subjects (DIS2026)\\
4-8 May 2026\\
Bologna, Italy\\}

\begin{document}
\maketitle

\ \\[-1.75cm]

%%%%%%%%%%%%%%%%%%%%%%%%%%%%%%%%%%%%%%%%%%%%%%%%%%%%%%%%%%%%%%%%%%%%%%%%%%%%%%%%
\section{Introduction}
\label{introduction}
\vspace{-0.15cm}

Nuclear modifications of structure functions have been investigated mainly
in the structure function $F_2^A$, where $A$ indicates a nucleus. 
The function $F_2^A$ contains both longitudinal and transverse components 
defined by the virtual-photon polarization vector in the charged-lepton scattering.
On the other hand, the structure function $F_1^A$ contains only
the transverse component. If both $F_1^A$ and $F_2^A$ are measured,
it is possible to investigate nuclear effects on 
both the longitudinal and transverse structure functions. 
%%%%%
So far, any nuclear effect is not found on the longitudinal-transverse 
structure function ratio $R_A = F_L^A/(2xF_1^A)$, where $F_L^A$ is the
longitudinal structure function. However, there were
discussions on this topic around 2000, when experimental groups started
reporting their data on $R_A$. There were reports
from the HERMES, CCFR/NuTeV, and JLab-Hall-C collaborations 
from 2000 to 2007 \cite{HERMES:1999bwb,CCFRNuTeV:2001njk,Tvaskis:2006tv},
and these measurements did not indicate any significant nuclear modification
in $R_A$ by considering their experimental errors. Based on these experimental 
measurements, it has been assumed in the analysis of lepton-nucleus scattering
data that there exist no nuclear effect in the longitudinal-transverse 
structure-function ratio.

However, as it was pointed out theoretically in 2003 \cite{Ericson:2002ep}
that nuclear effect should exist in the function $R_A$ by the standard
theoretical formalism due to nucleon's Fermi motion.
The longitudinal and transverse components of the structure functions
are defined by the virtual-photon polarization vector in charged-lepton
scattering by taking the virtual-photon momentum direction 
along the $z$ axis. If we consider the center-of-momentum frame
between the free nucleon and the virtual photon, the nucleon also
moves along the $z$ direction. However, a nucleon in a nucleus moves
in any direction, which is not necessary along the $z$ axis, so that
the longitudinal and transverse structure functions could mix 
within the nucleus due to the transverse Fermi motion.
Such mixing effects would be of the order of $\vec p_\perp^{\,2}/Q^2$, where
$\vec p_\perp$ is the transverse momentum of the nucleon in the nucleus
and $Q^2$ is defined by $Q^2=-q^2$ with the virtual photon momentum $q$.

Although it is obvious that nuclear modifications should exist
from the theoretical point of view, it is unfortunate that
this phenomenon has not been found experimentally and that
lepton-nucleus scattering data have been 
analyzed by assuming that such a nuclear modification of $R_A$ does not exist.
In determining parton distribution functions (PDFs) of the ``nucleon",
light-nucleus data are often included in global analyses without
worrying about the existence of such nuclear modifications.
In these days, the PDFs become more and more accurate with years
and they are used for finding new physics beyond the standard model
of particle physics. Therefore, it is desirable to have accurate PDFs
by including the nuclear modifications of $R_A$ in future analyses.
%%%%%
In addition, the nuclear effects could affect the other research
topics such as the short-range corrections (SRCs) \cite{SRC-JLab} 
as the phenomena appear in the large-$x$ region, 
where $x$ is the Bjorken scaling variable.
In the SRC experiments, used nuclei are from the deuteron 
to a large nucleus of lead ($^{208}$Pb).
Nuclear modifications should be large 
in the longitudinal-transverse ratio $R_A$ for large nuclei; however, 
they are neglected in the experimental analysis of the SRCs. 
Therefore, it could be important to consider such nuclear modifications.

Because the $R_A$ experiment is now in progress for the deuteron
at JLab \cite{R-Jlab-pro-R}
and the polarized deuteron experiment is also under preparation
at JLab \cite{Poudel:2025nof}, nuclear modifications are explained
for the function $R_A$ of the deuteron in this paper by quoting
the recent work \cite{Kumano:2025qzm}.
In Sec.\,\ref{formalism}, the theoretical formalism is explained
for calculating the nuclear modifications by using the convolution model.
Numerical results are shown in Sec.\,\ref{results} and
they are summarized in in Sec.\,\ref{summary}.

% \vfill\eject

%%%%%%%%%%%%%%%%%%%%%%%%%%%%%%%%%%%%%%%%%%%%%%%%%%%%%%%%%%%%%%%%%%%%%%%%%%%%%%%%
\section{Formalism}
\label{formalism}

The charged-lepton deep inelastic cross section from a nucleus ($A$)
or a nucleon ($N$) is expressed by the lepton tensor multiplied 
by the hadron tensor $W^{A,N}_{\mu\nu}$.
For studying the longitudinal and transverse components of
the structure functions, we write photon-helicity ($\lambda$) dependence 
of the hadron tensor by multiplying the photon polarization vector 
$\varepsilon_\lambda^{\,\mu}$ as
\begin{align}
W^{A,N}_\lambda (p_{_{A,N}}, q) 
            = \varepsilon_\lambda^{\,\mu *} \varepsilon_\lambda^{\nu} 
                  W^{A,N}_{\mu\nu} (p_{_{A,N}}, q) .
\label{eqn:W_lambda}
\end{align}
On the other hand, the hadron tensor $W^{A,N}_{\mu\nu}$ is generally expressed 
by two structure functions $W^{A,N}_1$ and $W^{A,N}_2$ as
\begin{align}
W^{A,N}_{\mu\nu} (p_{_{A,N}},  q)  =
  - W^{A,N}_1 (p_{_{A,N}}, q) 
  \left ( g_{\mu\nu} - \frac{q_\mu q_\nu}{q^2} \right )
+ W^{A,N}_2 (p_{_{A,N}}, q) \, \frac{\pt_{_{A,N} \mu} 
  \, \pt_{_{A,N} \nu}}{p_{_{A,N}}^2} .
\label{eqn:hadron}
\end{align}
Here, $\pt_{\mu}$ is defined by 
$\pt_{\mu} = p_{\mu} -(p \cdot q) \, q_\mu /q^2$,
and $p_{_{A,N}}$ is the nucleus or nucleon momentum.
%%%
The transverse and longitudinal structure functions 
$W^{A,N}_T$ and $W^{A,N}_L$
are defined by the helicity-dependent function $W^{A,N}_\lambda$,
 $W^{A,N}_{1}$, and $W^{A,N}_{2}$ as
\begin{align}
W^{A,N}_T = 
    \frac{ W^{A,N}_{\lambda=+1} + W^{A,N}_{\lambda=-1} }{2}
     = W^{A,N}_{1} ,
\ \ \ 
W^{A,N}_L = W^{A,N}_{\lambda=0}
     = \left ( 1 + \frac{\nu_{_{A,N}}^2}{Q^2} \right) W^{A,N}_2 - W^{A,N}_1 ,
\label{eqn:W^AN_TL}
\end{align}
where $\nu_N$ and $\nu_A$ are defined by
$ \nu_A = p_{_A} \cdot q / \sqrt{p_{_A}^{\,2}} = \nu$ and
$ \nu_N = p_{_N} \cdot q / \sqrt{p_{_N}^{\,2}}$ .
%%%%%%
Instead of $W_{1,2,T,L}$, the structure functions $F_{1,2,T,L}$ are usually
used nowadays in showing experimental data and theoretical calculations.
They are related to $W_{1,2,T,L}$ as
\begin{align}
F_1^{A,N} & = F_T^{A,N}  = \sqrt{p_{_{A,N}}^2} \, W_1^{A,N} ,
\nonumber \\
F_2^{A,N} & = \frac{p_{_{A,N}} \cdot q}{\sqrt{p_{_{A,N}}^2}} \, W_2^{A,N} ,
\nonumber \\
F_L^{A,N} & = \left ( 1 + \frac{Q^2}{\nu_{_{A,N}}^2} \right ) F_2^{A,N} 
            - 2 x_{_{A,N}} F_1^{A,N} .
\label{eqn:F^AN_TL}
\end{align}
The kinematical variables $x_{A,N}$, $x$, and $y$ are defined by
\begin{align}
x_A  = \frac{Q^2}{ 2 \, p_A \cdot q}, \ \ \ 
x_N  = \frac{Q^2}{ 2 \, p_N \cdot q} = \frac{x}{y}, \ \ \ 
x    = \frac{Q^2}{2 \, M_N \nu},  \ \ \ 
y    = \frac{p_N \cdot q}{M_N \, \nu},
\end{align}
where $M_N$ and $M_A$ are nucleon and nuclear masses, respectively.

In Eqs.\,(\ref{eqn:W^AN_TL}) and (\ref{eqn:F^AN_TL}), we find that 
$F_1^{A,N}$ are transverse components and $F_2^{A,N}$ contain 
both longitudinal and transverse ones. 
By combining these two structure functions
$F_1^{A,N}$ and $F_2^{A,N}$, the longitudinal structure functions 
$F_L^{A,N}$ are obtained. 
The longitudinal-transverse structure function ratio is then
given as
\begin{align}
R_{A,N} (x_A, Q^2) = \frac{F_L^{A,N} (x_{A,N}, Q^2)}
 {2 \, x_{A,N} F_1^{A,N} (x_{A,N}, Q^2)}  .
\label{eqn:RAN}
\end{align}
In the Bjorken scaling limit, the Callan-Gross relation 
$F_2^{A,N} = 2 x_{_{A,N}} F_1^{A,N}$
should be valid. It means that the functions $R_{A,N}$ should vanish
in this limit, so that the functions $R_{A,N}$ at finite $Q^2$
contain the effects of higher twist and higher order of 
the running coupling constant $\alpha_s$.
This fact suggests that the functions $R_{A,N}$ are appropriate
quantities to investigate dynamical aspects of nucleon or 
nuclear structure functions.

In the recent work of Ref.\,\cite{Kumano:2025qzm}, 
a standard convolution model is used
for calculating the nuclear structure functions.
It is given by the nucleon's hadron tensor $ W^N_{\mu\nu}$
convoluted with the nucleon's spectral function $S(\pn)$ as
\begin{align}
W^A_{\mu\nu} (\pa, q) 
  = {\displaystyle\int} d^4 \pn \, S(\pn) \, W^N_{\mu\nu} (\pn, q).
\label{eqn:WA}
\end{align}
This theoretical method is used as a standard way to calculate
nuclear structure functions from the smallest nucleus, the deuteron,
to large nuclei \cite{Hirai:2010xs,Cosyn:2017fbo,Kumano:2026ipc}.
A simple spectral function is calculated by 
the momentum-space wave function $\phi_i$,
where $i$ is the $i$-th nucleon in the nucleus,
as
\begin{align}
S (p_N)  = \sum_i | \phi _i (\vec p_N) |^2  \delta 
   \left ( p_N^{\, 0} - M_A + \sqrt{M_{A-i}^{\ 2} 
          +\vec p_N^{\ 2}} \, \right ),
\label{eqn:spectral}
\end{align}
where the residual nuclear mass $M_{A-i}$ is given by 
the separation energy $\varepsilon_i$ 
by $\varepsilon_i = (M_{A-i}+M_N) - M_A$.
The average separation energy is taken as
$\varepsilon_i \rightarrow \left< \varepsilon \right>$
and the non-relativistic kinematic is taken 
for the squared root in Eq.\,(\ref{eqn:spectral}).
Then, $p_N^0$ and $\left< \varepsilon \right>$ are related by
$ p_N^{\, 0} = M_N - \left< \varepsilon \right> 
   - \vec p_N^{\ 2} / (2 M_{A-1})$ with the replacement
$M_{A-i} \to M_{A-1}$.

In order to extract the structure functions
$W_{1,2}^A$ or $F_{1,2,L}$ from $W^A_{\mu\nu}$ in 
Eq.\,(\ref{eqn:WA}), their projection operators \cite{Kimura:2008gz}
are multiplied.
Then, the nuclear structure functions are written
in the convolution forms as
\begin{align}
\left(
    \begin{aligned}
      \,      F_2^A(x_A,Q^2) \, \\
      \, 2x_A F_1^A(x_A,Q^2) \, \\
      \,      F_L^A(x_A,Q^2) \,
    \end{aligned}
\right)
& \, = \int_x^A dy \,
\left(
    \begin{aligned}
      \, f_{22} (y) \, & \ \ \ \, 0 \,      & \, 0 \ \ \ \,  \\
      \, 0 \ \ \ \     & \, f_{11} (y) \, & \, f_{1L} (y) \,\\
      \, 0 \ \ \ \    & \, f_{L1} (y) \, & \, f_{LL} (y) \,
    \end{aligned}
\right) \,
\left(
    \begin{aligned}
      \,               F_2^N(x/y,Q^2) \, \\
      \, 2 \, \frac{x}{y} F_1^N(x/y,Q^2) \, \\
      \,               F_L^N(x/y,Q^2) \,
    \end{aligned}
\right) .
\label{eqn:12L-convolution}
\end{align}
The lightcone momentum distributions of the nucleon
$f_{22}$, $f_{LL}$, $f_{11}$, $f_{L1}$, and $f_{1L}$
are given by
\begin{align}
f_{22} (y) & = \int_0^\infty dp_{N\perp} 2 \pi p_{N\perp} y
             \frac{M_N \nu}{|\vec q \, |} 
             \left | \phi (\vec p_N) \right |^2
  \left [ \frac{2 \, (x/y)^2 \, \vec p_{N\perp}^{\ 2}}{(1+Q^2/\nu^2)Q^2}
 + \left ( 1 + \frac{2 \, (x/y) \, 
        p_{N\parallel}}{\sqrt{Q^2+\nu^2}} \right )^2 \right ],
\nonumber \\
f_{LL} (y) & = f_{11} (y) 
    = \int_0^\infty dp_{N\perp} 2 \pi p_{N\perp} y
             \frac{M_N \nu}{|\vec q \, |} 
             \left | \phi (\vec p_N) \right |^2
    \left ( 1 + \frac{\vec p_{N\perp}^{\ 2}}{\pt_N^{\,\, 2}} \right ) ,
\nonumber \\
f_{L1} (y) & = f_{1L} (y)
    = \int_0^\infty dp_{N\perp} 2 \pi p_{N\perp} y
             \frac{M_N \nu}{|\vec q \, |} 
             \left | \phi (\vec p_N) \right |^2
             \frac{\vec p_{N\perp}^{\ 2}}{\pt_N^{\,\, 2}} .
\label{eqn:fL1}
\end{align}
The longitudinal-trasverse mixture appears as
the functions $f_{L1} (y)$ and $f_{1L} (y)$, which are proportional
to $\vec p_{N\perp}^{\ 2} / \pt_N^{\,\, 2}$. 
This mixture coefficient is given as 
\begin{align}
\frac{\vec p_{N\perp}^{\ 2}}{\pt_N^{\,\, 2}}
= \frac{4 x_N^2 \vec p_{N\perp}^{\ 2}}
       {\left( 1+\frac{4 x_N^2 p_N^2}{Q^2} \right)Q^2}
\approx \frac{4 x_N^2 \vec p_{N\perp}^{\ 2}}
       {\left( 1+\frac{4 x_N^2 M_N^2}{Q^2} \right)Q^2}
\sim O \left( \frac{\vec p_{N\perp}^{\ 2}}{Q^2} \right) .
\label{eqn:ptN2-Q2}
\end{align}
In Eq.\,(\ref{eqn:12L-convolution}), the structure function $F_2^N$ does not mix;
however, $F_1^N$ and $F_L^N$ mix with the mixing coefficient
proportional to $\vec p_{N\perp}^{\ 2}/Q^2$.
The nuclear modifications due to this mixture should be apparent
at small $Q^2$.
The existence of the transverse Fermi motion of a nucleon in a nucleus
makes it possible to mix the longitudinal and transverse structure functions.

%%%%%%%%%%%%%%%%%%%%%%%%%%%%%%%%%%%%%%%%%%%%%%%%%%%%%%%%%%%%%%%%%%%%%%%%%%%%%%%%
\section{Results}
\label{results}

The nuclear structure functions $F_{1,2,L}^A$ in Eqs.\,(\ref{eqn:12L-convolution}) 
and (\ref{eqn:fL1}) are calculated numerically if the momentum-space
wave function $\phi (\vec p_N)$ and the separation energy 
$\left< \varepsilon \right>$ are supplied for the deuteron.
In addition, the unpolarized parton PDFs and the longitudinal-transverse ratio
$R_N$ for the nucleon should be given in order to have
$F_{1,2,L}^N$ for the nucleon.
The function $F_{2}^N$ is calculated by the leading-order PDFs
of MSTW08 in $\alpha_s$, and then $F_{1}^N$
is calculated by this function and 
the SLAC 1990 parametrization of $R_N$ as 
\begin{align}
 F_1^N (x_N,Q^2) = \frac{ ( 1+Q^2/\nu_N^2 ) \, 
 F_2^N (x_N,Q^2)}{2 \, x_N \, \{ 1+R_N (x_N,Q^2) \} }.
 \end{align}
The experimental separation energy of the deuteron is 2.22457 MeV,
and the Bonn wave function is used for the deuteron.

%%%%%%%%%%%%%%%%%%%%%%%%%%%%%%%%%%%%%%%%%%%%%%%%%%%%%%%%%%%%%%%
\begin{figure}[b]
\vspace{-0.30cm}
\begin{minipage}[c]{0.47\textwidth}
     \hspace{-0.20cm}
     \includegraphics[width=7.7cm]{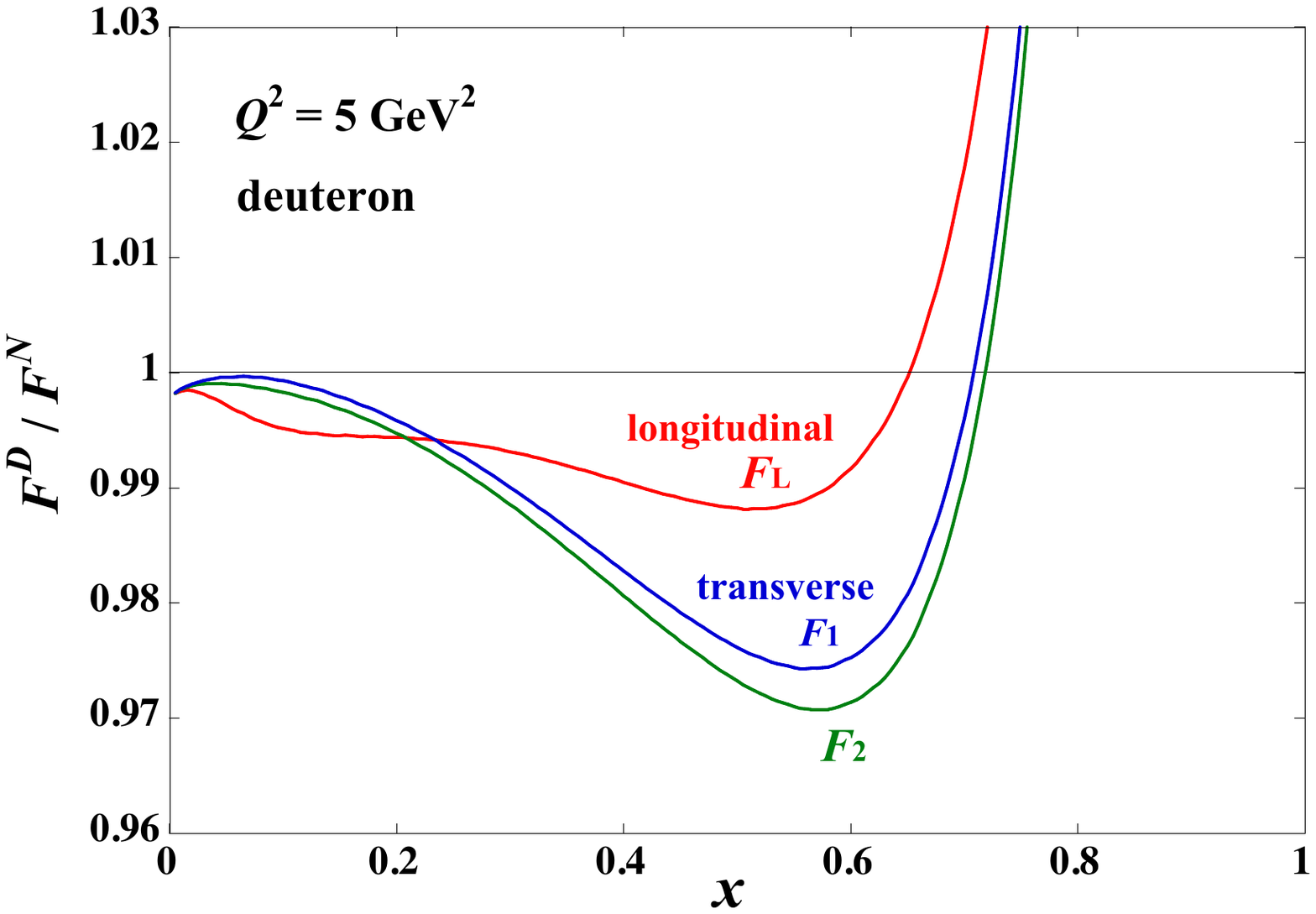}
\vspace{-0.90cm}
\caption{Nuclear modifications of the structure functions
$F_1$, $F_2$, and $F_L$ at $Q^2 =5$ GeV$^2$ in the deuteron.}
\label{fig:F2F1FL}
\vspace{-0.4cm}
\end{minipage}
%%%%%%%%%%
\hspace{0.50cm}\vspace{-0.30cm}
%%%%%%%%%%
\begin{minipage}[c]{0.47\textwidth}
     \hspace{-0.20cm} \vspace{-0.05cm}
     \includegraphics[width=7.7cm]{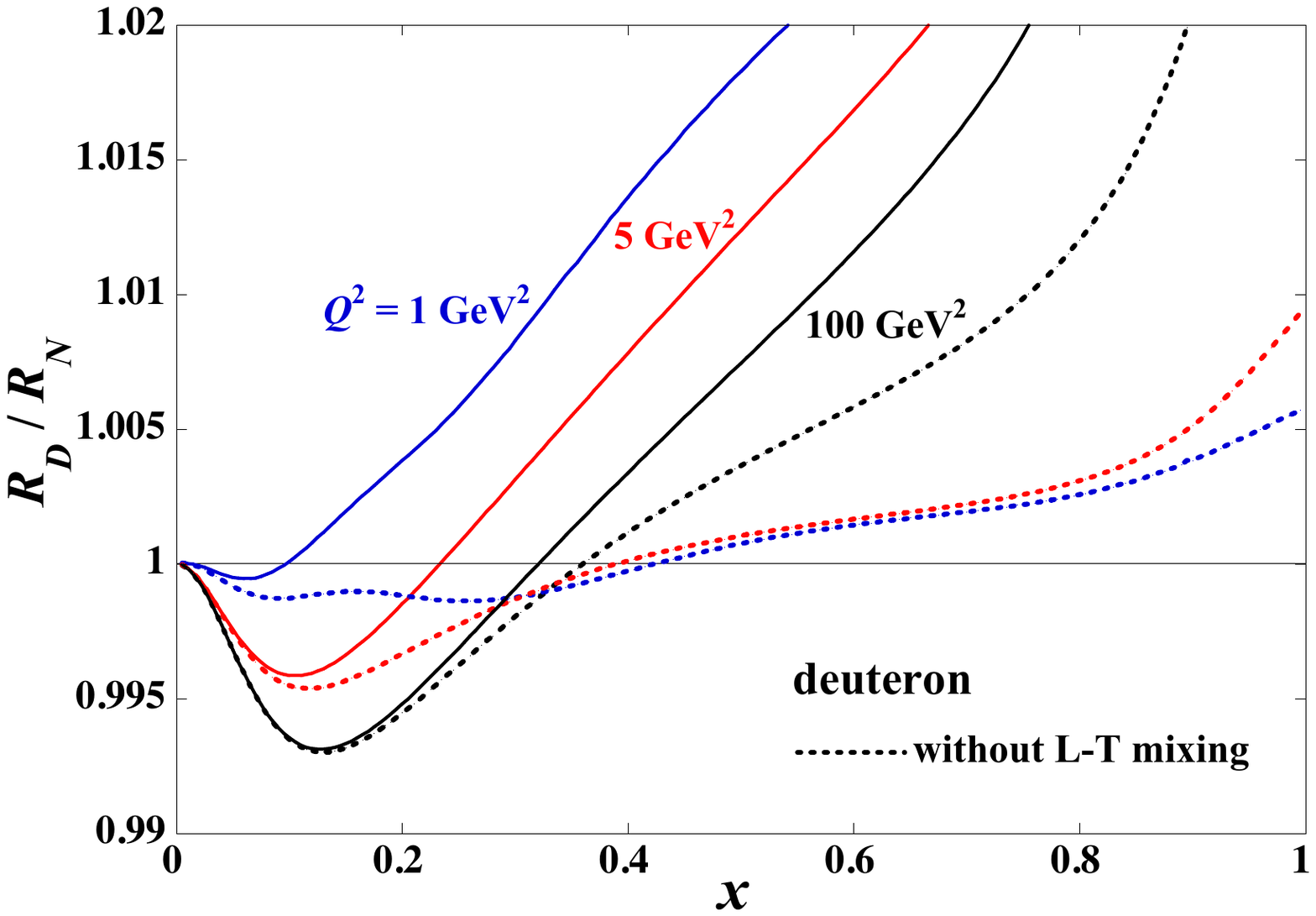}
\vspace{-0.90cm}
\caption{Nuclear modifications of 
$R_N$ for the deuteron 
at $Q^2=1$, 10, and 100 GeV$^2$.}
\label{fig:R-1-10-100}
\vspace{-0.0cm}
\end{minipage}
\end{figure}
%%%%%%%%%%%%%%%%%%%%%%%%%%%%%%%%%%%%%%%%%%%%%%%%%%%%%%%%%%%%%%%

Calculated nuclear modifications are shown as the ratios
$F_{1}^D/F_{1}^N$, $F_{2}^D/F_{2}^N$, and $F_{L}^D/F_{L}^N$ in
Fig.\,\ref{fig:F2F1FL} at $Q^2 = 5$ GeV$^2$.
In this convolution formalism, nuclear-binding and Fermi-motion
effects are included in the spectral function. Their contributions
have been investigated mainly in $F_2^A$, and their $x$ dependencies
are typically shown in this figure.
The nuclear binding contributes to the negative modifications 
at medium $x$ ($\sim 0.4$), and the the steep positive increase
at large $x$ ($\sim 0.7$) is due to the nucleon's Fermi motion.
In general, the nuclear modifications are within the order of
a few percent as the deuteron is the smallest nucleus.
%%%%%
In Fig.\,\ref{fig:F2F1FL}, the nuclear modifications are also
shown separately for the longitudinal ones by $F_{L}^D/F_{L}^N$
and the transverse ones by $F_{1}^D/F_{1}^N$,
because $F_2^A$ has both longitudinal and transverse components. 
The nuclear modifications of the transverse component
$F_1^N$ are similar to the ones of $F_2^N$; however,
the longitudinal modifications are very different 
from the ones of $F_2^N$ and $F_1^N$.

Next, the nuclear modifications are shown by the solid curves
for the longitudinal-transverse ratio by $R_D/R_N$ 
at $Q^2=1$, 10, and 100 GeV$^2$ in Fig.\,\ref{fig:R-1-10-100}.
The dashed curves are obtained by terminating the mixing terms,
$\vec p_{N\perp}^{\ 2} / \pt_N^{\,\, 2} \to 0$
in $f_{LL} (y)$, $f_{11} (y)$, $f_{L1} (y)$, and $f_{1L} (y)$,
and also by taking $[ \cdot\cdot\cdot ] \to 1$ in $f_{2} (y)$.
Even though the longitudinal-transverse mixing does not exist,
there are finite modifications even at $Q^2=100$ GeV$^2$.
There are two major sources for the nuclear effects:
\vspace{-0.20cm}
\begin{itemize}
\setlength{\itemsep}{-0.10cm}
\item[(1)] One is the longitudinal-transverse mixing effect as shown
by the differences between the solid and dashed curves.
\item[(2)] The other is the nuclear modifications due to the difference
between the $x$-dependent functional forms of $F_L^N$
and $2 x_N F_1$. This difference appears as
the nuclear modifications though the convolution integrals
with the nucleon momentum distributions.
\vspace{-0.20cm}
\end{itemize}
Although the first mixing effects disappear with increasing $Q^2$
as $\vec p_{N\perp}^{\ 2}/Q^2 \to 0$, there are relatively large
modifications even at $Q^2=100$ GeV$^2$ due to the second factor.
Because the second nuclear effects are not large at $Q^2=5$ GeV$^2$
in Fig.\,\ref{fig:R-1-10-100}, the large modification differences
in $F_{L}^D/F_{L}^N$ from $F_{1}^D/F_{1}^N$ and $F_{2}^D/F_{2}^N$ 
at $Q^2=5$ GeV$^2$ in Fig.\,\ref{fig:F2F1FL} should come from 
the first mixing effects.

I comment on related works and other corrections at large $x$.
The nuclear modifications of $F_2^A$ were studied
for the deuteron, $^3$He, and $^3$H in Ref.\,\cite{Pace:2001cm},
and they are also studied
by including target-mass corrections (TMCs), off-shell effects,
and pion contributions in Ref.\,\cite{Kulagin:2004ie}.
In comparison with the actual experimental data of $R_A$,
these other corrections need to be taken into account \cite{Tropiano:2018quk}.
The TMCs were shown numerically in Ref.\,\cite{Brady:2011uy},
by using the operator product expansion,
the collinear factorization of Ellis, Furmanski, and Petronzio,
$\xi$ scaling, and Accardi and Qiu method.
For the deuteron, the large Fermi-motion effects could be
reduced by these TMCs depending on the prescription.
Typical effects were shown in Fig.\,3 of Ref.\,\cite{Kumano:2025qzm}
by  taking the $\xi$ scaling as an example.
The results suggest that the TMCs suppress
the Fermi-motion rise and such effects are large 
at small $Q^2$ ($=1$ GeV$^2$) due to 
the $M_N^2/Q^2$ factor.

Since a simple convolution model is used in this work to calculate
the deuteron's structure functions, the physics of small $x$ ($x<0.1$)
is not considered, and this region should be explained by a theoretical
model which can explain the nuclear shadowing.
Because the higher-twist and higher-order effects from the gluon are 
essential in this region, the modifications
of $R_N$ at small $x$ should become interesting topic, especially
in connection with the gluon saturation. We need further research 
progress in this region both theoretically and experimentally.

In this work, it is shown theoretically that the nuclear modifications
exist in the function $R_N$, whereas such an existence has been
neglected in analyzing nuclear data.
Since the experiment is under investigation at JLab for 
the function $R_A$ of the deuteron, we expect that 
such modification effects could become clearer.
However, because the nuclear effects are small in the deuteron,
it would be desirable to investigate such effects
in large nuclei. We should note that no nuclear modification of $R_N$
has been considered in analyzing heavy-nuclear data in lepton scattering.
For example, such nuclear effects are totally neglected in the SRC studies,
although large nuclei are used and the SRC effects appear at large $x$.
In addition, global analyses have been done without the
nuclear effects in $R_A$ for determining the nucleon's PDFs,
although light nuclear data are included in the analyses.
Furthermore, the small-$x$ part could probe
interesting gluon dynamics, which is a different in physics
from the nuclear binding and Fermi motion in this work.

\vfill\eject

%%%%%%%%%%%%%%%%%%%%%%%%%%%%%%%%%%%%%%%%%%%%%%%%%%%%%%%%%%%%%%%%%%%%%%%%%%%%%%%%
\section{Summary}
\label{summary}

Nuclear modifications were investigated for the longitudinal-transverse
structure-function ratio $R_A$ for the deuteron by using a standard 
convolution model. The nuclear modifications are vey different for
the longitudinal function $F_L^A$ from the ones for 
the functions $F_1^A$ and $F_2^A$. It leads to significant
nuclear effects in the longitudinal-transverse ratio $R_A$.
So far, there is no experimental evidence for the nuclear effect
in any nuclear $R_A$, so that it is assumed that the nuclear effects
do not exist in $R_A$ in analyzing lepton scattering experimental data
from nuclei. As shown in this work, it is not appropriate 
for precise analyses of experimental measurements 
and in extracting physical consequences from lepton scattering data.
Hopefully, the nuclear modifications in $R_A$ will be investigated
in future both theoretically and experimentally to establish
the field of nuclear effects in the structure functions
and to provide accurate information on the PDFs and also
short-range correlation phenomena, where no nuclear modification
is applied in analyzing even heavy-nuclear data.

%%%%%%%%%%%%%%%%%%%%%%%%%%%%%%%%%%%%%%%%%%%%%%%%%%%%%%%%%%%%%%%%%%%%%%%%%%%%%%%%

%%%%%%%%%%%%%%%%%%%%%%%%%%%%%%%%%%%%%%%%%%%%%%%%%%%%%%%%%%%%%%%%%%%%%%%%%%%%%%%%


\begin{thebibliography}{99}
\setlength{\itemsep}{-0.01cm}
\setlength\baselineskip{14pt}
\vspace{-0.22cm}
\bibitem{HERMES:1999bwb}
    K. Ackerstaff et al. (HERMES Collaboration),
%    \emph{Nuclear effects on R = $\sigma_L$ / $\sigma_T$ 
%         in deep inelastic scattering},
    \href{https://doi.org/10.1016/S0370-2693(99)01493-8}
    {\emph{Phys. Lett. B} {\bf 475} (2000) 386};
%    [\href{https://arxiv.org/abs/hep-ex/9910071}{\tt arXiv:hep-ex/9910071}]
%%%%    \inspire{HERMES:1999bwb}.
%    [\href{https://inspirehep.net/literature/509396}{inSPIRE}].
     A. Airapetian et al.,
    \href{https://doi.org/10.1016/j.physletb.2003.06.044}
    {\emph{Erratum: Phys. Lett. B} {\bf 567} (2003) 339}.
%    [\href{https://arxiv.org/abs/hep-ex/0210067}{\tt arXiv:hep-ex/0210067}]
%%%%    \inspire{HERMES:1999bwb}.
%    [\href{https://inspirehep.net/literature/509396}{inSPIRE}].
\bibitem{CCFRNuTeV:2001njk}
    U. K. Yang et al. (CCFR/NuTeV Collaboration),
%    \emph{Extraction of R = sigma(L) / sigma(T) from CCFR Fe-neutrino(muon) 
%          and Fe-anti-neutrino(muon) differential cross-sections},
    \href{https://doi.org/10.1103/PhysRevLett.87.251802}
    {\emph{Phys. Rev. Lett.} {\bf 87} (2001) 251802}.
%    [\href{https://arxiv.org/abs/hep-ex/0104040}{\tt arXiv:hep-ex/0104040}]
%%%%    \inspire{CCFRNuTeV:2001njk}.
%    [\href{https://inspirehep.net/literature/555716}{inSPIRE}].
\bibitem{Tvaskis:2006tv}
    V. Tvaskis et al.,
%    \emph{Longitudinal-transverse separations of structure functions 
%          at low Q**2 for hydrogen and deuterium},
    \href{https://doi.org/10.1103/PhysRevLett.98.142301}
    {\emph{Phys. Rev. Lett.} {\bf 98} (2007) 2003}.
%    [\href{https://arxiv.org/abs/nucl-ex/0611023}{\tt arXiv:nucl-ex/0611023}]
%%%%    \inspire{Tvaskis:2006tv}.
%    [\href{https://inspirehep.net/literature/731790}{inSPIRE}].
\bibitem{Ericson:2002ep}
    M. Ericson and S. Kumano,
%    \emph{Nuclear modification of transverse longitudinal structure function ratio},
    \href{https://doi.org/10.1103/PhysRevC.67.022201}
    {\emph{Phys. Rev. C} {\bf 67} (2003) 022201}.
%    [\href{https://arxiv.org/abs/hep-ph/0212001}{\tt arXiv:hep-ph/0212001}]
%%%%    \inspire{Ericson:2002ep}.
%    [\href{https://inspirehep.net/literature/603425}{inSPIRE}].
\bibitem{SRC-JLab} 
     B. Schmookler {\it et al.} (CLAS Collaboration),
     \href{https://www.nature.com/articles/s41586-019-0925-9}
     {\emph{Nature} {\bf 566} (2019) 354};
     A. W. Denniston {\it et al.},
     \href{https://doi.org/10.1103/PhysRevLett.133.152502}
     {\emph{Phys. Rev. Lett.} {\bf 133} (2024) 152502};
     Top 10 breakthoughs of 2024 in physics, Physics World, 
     https://physicsworld.com/a/top-10-breakthroughs-of-the-year-in-physics-for-2024-revealed/.
\bibitem{R-Jlab-pro-R} 
     R. Ent {\it et al.}, 
     \href{https://www.jlab.org/exp_prog/proposals/proposal_updates/PR12-06-104_pac36.pdf}
     {Update of JLab proposal E12-06-104 (2010)}.
\bibitem{Poudel:2025nof}
%    Jiwan Poudel, Alessandro Bacchetta, Jian-Ping Chen, 
%    and Nathaly Santiesteban, 
    J. Poudel, A. Bacchetta, J.-P. Chen, 
    and N. Santiesteban, 
%    \emph{Experimental study of tensor structure function of deuteron},
    \href{https://doi.org/10.1140/epja/s10050-025-01558-w}
    {\emph{Euro. Phys. J. A} {\bf 61} (2025) 81}.
%    [\href{https://arxiv.org/abs/2506.04506}{\tt arXiv:2506.04506}]
%%%%    \inspire{Poudel:2025nof}.
%    [\href{https://inspirehep.net/literature/1799442}{inSPIRE}].
\bibitem{Kumano:2025qzm}
    S. Kumano,
%    \emph{Existence of nuclear modifications of the nucleon 
%         longitudinal-transverse structure-function ratio},
    \href{https://doi.org/10.1103/xgbh-grqx}
    {\emph{Phys. Rev. C} {\bf 113} (2026) 015206}.
%    \href{https://arxiv.org/abs/2506.18305}{\tt arXiv:2506.18305}
%%%%    \inspire{Kumano:2025qzm}.
%    [\href{https://inspirehep.net/literature/2937835}{inSPIRE}].
\bibitem{Hirai:2010xs}
    M. Hirai, S. Kumano, K. Saito, and T. Watanabe,
%    \emph{Clustering aspects in nuclear structure functions},
    \href{https://doi.org/10.1103/PhysRevC.83.035202}
    {\emph{Phys. Rev. C} {\bf 83} (2011) 035202}.
%    [\href{https://arxiv.org/abs/1008.1313}{\tt arXiv:1008.1313}]
%%%%    \inspire{Hirai:2010xs}.
%    [https://inspirehep.net/literature/864843}{inSPIRE}].
\bibitem{Cosyn:2017fbo}
    W. Cosyn, Yu-Bing Dong, S. Kumano, and M. Sargsian,
%       \emph{Tensor-polarized structure function $b_1$
%             in the standard convolution description of the deuteron},
       \href{https://doi.org/10.1103/PhysRevD.95.074036}
       {\emph{Phys. Rev. D} {\bf 95} (2017) 074036}. 
%%%%%    \inspire{Cosyn:2017fbo}. 
%    [\href{https://inspirehep.net/literature/1514035}{inSPIRE}].
\bibitem{Kumano:2026ipc}
    S. Kumano and K. Kuroki,
%       \emph{Tensor-polarized parton distribution functions 
%             of the deuteron by a convolution model},
       \href{https://arxiv.org/abs/2607.09237}
       {arXiv:2607.09237 (2026)}. 
%%%%%    \inspire{Kumano:2026ipc}. 
%    [\href{https://inspirehep.net/literature/3179588}{inSPIRE}].
\bibitem{Kimura:2008gz}
    T.-Y. Kimura and S. Kumano,
%       \emph{Projections of structure functions in a spin-one hadrons},
       \href{https://doi.org/10.1103/PhysRevD.78.117505}
       {\emph{Phys. Rev. D} {\bf 78} (2008) 117505}.
%%%%%    \inspire{Kimura:2008gz}. 
%    [\href{https://inspirehep.net/literature/801956}{inSPIRE}].
\bibitem{Pace:2001cm}
     E. Pace, G. Salme, S. Scopetta, and A. Kievsky,
%       \emph{Neutron structure function F(2)**n (x) from deep inelastic 
%             electron scattering off few nucleon systems},
       \href{https://doi.org/10.1103/PhysRevC.64.055203}
       {\emph{Phys. Rev. C} {\bf 64} (2001) 055203}.
%%%%%    \inspire{Pace:2001cm}. 
%    [\href{https://inspirehep.net/literature/562321}{inSPIRE}].
\bibitem{Kulagin:2004ie}
     S. A. Kulagin and R. Petti, 
%       \emph{Global study of nuclear structure functions},
       \href{https://doi.org/10.1016/j.nuclphysa.2005.10.011}
       {\emph{Nucl. Phys. A} {\bf 765} (2006) 126}.
%%%%%    \inspire{Kulagin:2004ie}. 
%    [\href{https://inspirehep.net/literature/668239}{inSPIRE}].
\bibitem{Tropiano:2018quk}
     A. J. Tropiano, J. J. Ethier, W. Melnitchouk, and N. Sato,
%       \emph{Deep-inelastic and quasielastic electron scattering from $A=3$ nuclei},
       \href{https://doi.org/10.1103/PhysRevC.99.035201}
       {\emph{Phys. Rev. C} {\bf 99} (2006) 035201}.
%%%%%    \inspire{Tropiano:2018quk}. 
%    [\href{https://inspirehep.net/literature/1704292}{inSPIRE}].
\bibitem{Brady:2011uy}
     L. T. Brady, A. Accardi, T. J. Hobbs, and W. Melnitchouk,
%       \emph{Next-to leading order analysis of target mass corrections 
%             to structure functions and asymmetries},
       \href{https://doi.org/10.1103/PhysRevD.84.074008}
       {\emph{Phys. Rev. D} {\bf 84} (2011) 074008};
       Erratum,
       \href{https://doi.org/10.1103/PhysRevD.85.039902}
       {\emph{Phys. Rev. D} {\bf 85} (2012) 039902}.
%%%%%    \inspire{Brady:2011uy}. 
%    [\href{https://inspirehep.net/literature/924839}{inSPIRE}].
\end{thebibliography}
\end{document}